\documentstyle[prl,aps]{revtex}
\begin{document}
\draft
\title{Efficient many-party controlled teleportation of multi-qubit quantum
information via entanglement}
\author{Chui-Ping Yang$^{1,2}$, Shih-I Chu$^{2}$, and Siyuan Han$^{1}$}
\address{$^1$Department of Physics and Astronomy, University of Kansas, Lawrence,\\
Kansas 66045}
\address{$^2$Department of Chemistry, University of Kansas, and Kansas Center\\
for Advanced Scientific Computing, Lawrence, Kansas 66045}
\maketitle

\begin{abstract}
We present a way to teleport multi-qubit quantum information from a sender
to a distant receiver via the control of many agents in a network. We show
that the original state of each qubit can be restored by the receiver as
long as all the agents collaborate. However, even if one agent does not
cooperate, the receiver can not fully recover the original state of each
qubit. The method operates essentially through entangling quantum
information during teleportation, in such a way that the required auxiliary
qubit resources, local operation, and classical communication are
considerably reduced for the present purpose.
\end{abstract}

\date{\today }


\begin{center}
{\bf I. INTRODUCTION AND MOTIVATION}
\end{center}

Over the past decade, scientists have made dramatic progress in the field of
quantum teleportation. Theoretically, since the work of Bennett {\it et. al.}
[1] on teleporting a qubit of unknown information with the aid of
Einstein-Podolsky-Rosen (EPR) correlation; quantum teleportation has been
extended from discrete-variable systems to continuous-variable systems [2-5]
and also from a single qubit to multi qubits [6-9]. On the other hand,
recent experiments have demonstrated quantum teleportation with photon
polarized states [10], optical coherent states [11], and nuclear magnetic
resonance [12].

In 1998, Karlsson and Bourennane [13] generalized Bennett's idea, by the use
of a three-qubit entangled Greenberger-Horne-Zeilinger (GHZ) state $\left|
000\right\rangle +\left| 111\right\rangle $ instead of an EPR pair. In their
work, they showed that an arbitrary unknown state of a qubit could be
teleported to either one of two receivers. But only one of the two (either
one) can fully reconstruct the qubit state conditioned on the measurement
outcome of the other. Since that work, Hillery {\it et. al.} [14] first
proposed the idea of quantum secret sharing, i.e., splitting a message into
several parts so that no subset of parts is sufficient to read the message,
but the entire set is. In their work, they showed how a qubit of information
can be secretly shared by two agents via a three-qubit GHZ state, and also
generalized this procedure to split a qubit of information among more than
two agents, through a four-qubit GHZ state. More recently, a number of works
on quantum secret sharing have also been proposed [15-20].

In this paper, we restrict ourselves to an issue, i.e., teleporting
multi-qubit information from a sender to a distant receiver via the control
of many agents in a network. We wish that the receiver can successfully get
access to the original state of each qubit, as long as all the agents
collaborate through local operation and classical communication. However,
even if one agent does not cooperate, the original state of each qubit can
not fully be recovered by the receiver. The topic here might be of
particular interest, since controlled teleportation is useful in networked
quantum information processing and cryptographic conferencing [21,22].

One possible approach is to use the method described in Refs. [13,14]
directly for the present task. To explore the feasibility of this approach,
let us first give a brief review on the procedure introduced in [13,14]. To
teleport an arbitrary unknown state $\left| \psi \right\rangle _A=\alpha
\left| 0\right\rangle _A+\beta \left| 1\right\rangle _A$ of one qubit $A$
from Alice to a receiver Bob via the control of $n$ agents in a network, the
method in [13,14] requires a ($n+2)$-qubit GHZ state $\left|
GHZ\right\rangle =\left| 0\right\rangle _a\left| 0\right\rangle _b\left|
0\right\rangle ^{\otimes n}+\left| 1\right\rangle _a\left| 1\right\rangle
_b\left| 1\right\rangle ^{\otimes n}$ shared by Alice, Bob, and the $n$
agents. Here, GHZ qubits $a$ and $b$ belong to Alice and Bob, respectively;
while the other $n$ GHZ qubits belong to the $n$ agents. The initial state $%
\left| \psi \right\rangle _A\otimes \left| GHZ\right\rangle $ for the whole
system can be rewritten as 
\begin{eqnarray}
&&\ \ \ \ \ \left| \Phi ^{+}\right\rangle _{Aa}(\alpha \left| 0\right\rangle
_b\left| 0\right\rangle ^{\otimes n}+\beta \left| 1\right\rangle _b\left|
1\right\rangle ^{\otimes n})+\left| \Phi ^{-}\right\rangle _{Aa}(\alpha
\left| 0\right\rangle _b\left| 0\right\rangle ^{\otimes n}-\beta \left|
1\right\rangle _b\left| 1\right\rangle ^{\otimes n})  \nonumber \\
&&\ +\left| \psi ^{+}\right\rangle _{Aa}(\alpha \left| 1\right\rangle
_b\left| 1\right\rangle ^{\otimes n}+\beta \left| 0\right\rangle _b\left|
0\right\rangle ^{\otimes n})+\left| \psi ^{-}\right\rangle _{Aa}(\alpha
\left| 1\right\rangle _b\left| 1\right\rangle ^{\otimes n}-\beta \left|
0\right\rangle _b\left| 0\right\rangle ^{\otimes n}),
\end{eqnarray}
where $\left| \Phi ^{\pm }\right\rangle _{Aa}=$ $\left| 00\right\rangle
_{Aa}\pm \left| 11\right\rangle _{Aa}$ and $\left| \psi ^{\pm }\right\rangle
_{Aa}=$ $\left| 01\right\rangle _{Aa}\pm \left| 10\right\rangle _{Aa}$ are
Bell states for the two qubits $A$ and $a.$ Based on (1), it can be shown
that:

(i) For every outcome of Alice's Bell-state measurement on her qubits $A$
and $a$, Bob can restore the original state $\alpha \left| 0\right\rangle
_A+\beta \left| 1\right\rangle _A$ of the message qubit $A$ through his
qubit $b$, provided that each agent cooperates with him. That is, each agent
performs a Hadamard transformation $\left| 0\right\rangle \rightarrow \left|
0\right\rangle +\left| 1\right\rangle $ and $\left| 1\right\rangle
\rightarrow \left| 0\right\rangle -\left| 1\right\rangle $ on his/her GHZ
qubit, then makes a measurement on his/her GHZ qubit in a single-qubit
computational basis $\left| 0\right\rangle $ and $\left| 1\right\rangle ,$ 
{\it a measurement basis in which all single-qubit measurements discussed
throughout this paper will be performed}, and finally sends his/her
measurement result (one-bit classical message) to Bob.

(ii) On the other hand, note that after Alice's Bell-state measurement, the $%
n+1$ parties (the $n$ agents plus Bob) are left sharing a ($n+1$)-qubit
state of the form $\alpha \left| 0\right\rangle _b\left| 0\right\rangle
^{\otimes n}\pm \beta \left| 1\right\rangle _b\left| 1\right\rangle
^{\otimes n}$ or $\alpha \left| 1\right\rangle _b\left| 1\right\rangle
^{\otimes n}\pm \beta \left| 0\right\rangle _b\left| 0\right\rangle
^{\otimes n}$ (depending on Alice's Bell-state measurement outcome). Thus,
even if one agent does not cooperate with Bob, the resulting density
operator for the qubit $b$ belonging to Bob would be $\rho _b=\left| \alpha
\right| ^2\left| 0\right\rangle _b\left\langle 0\right| +\left| \beta
\right| ^2\left| 1\right\rangle _b\left\langle 1\right| $ or $\left| \alpha
\right| ^2\left| 1\right\rangle _b\left\langle 1\right| +\left| \beta
\right| ^2\left| 0\right\rangle _b\left\langle 0\right| ,$ which implies
that Bob has amplitude information about Alice's message qubit $A$ but knows
nothing about its phase. Therefore, Bob can not fully gain the original
information of Alice's qubit $A$ even if one agent does not collaborate with
him.

Note that the above argument is for the controlled teleportation of
one-qubit information. The procedure to extend the method in [13,14] to
teleport $m$-qubit information from a sender to a distant receiver via the
control of $n$ agents is as follows. Firstly, the sender needs to prepare $m$
copies of a ($n+2$)-qubit GHZ state, and then sends each agent one GHZ qubit
for every ($n+2$)-qubit GHZ state, i.e., a total of $m$ GHZ qubits are
needed to be distributed to each agent. Secondly, each agent needs to
perform a Hadamard transformation and then measurement on each of his/her $m$
GHZ qubits, i.e., a total of $m$ single-qubit Hadamard transformations and $%
m $ single-qubit measurements are required for each agent. Lastly, each
agent needs to send the receiver all of his/her measurement results (i.e., $%
m $-bit classical message), in order for the receiver to restore the
original $m$-qubit information. Hence, the method in [13,14] requires
considerably large auxiliary qubit resources, local operation, as well as
classical communication, especially when the number of
``teleported''-message qubits is significantly large.

In the following, we describe a new way to implement the present task. The
method works actually by entangling quantum information during
teleportation. As shown below, regardless of the amount of information to be
teleported, the proposed approach only requires that: (i) the sender assigns 
{\it one qubit} to each agent; (ii) each agent performs {\it one
single-qubit Hadamard transformation} and {\it one single-qubit measurement}
on his/her qubit; and (iii) each agent sends {\it one-bit classical message}
to the receiver. Therefore, compared with the method in [13,14], the present
scheme is much simpler and economical, because the required auxiliary qubit
resources, the number of local operations, and the quantity of classical
communication are greatly reduced.

The paper is organized as follows. In Sec. II, we present a way to teleport
multi-qubit information to a distant receiver via the control of one agent.
We then make a comparison between our method and the method in [13,14]. In
Sec. III, we discuss how to decompose multi-qubit GHZ states and then
generalize our method to a multi-agent controlled teleportation. In Sec. IV,
we further apply the method to teleporting multiple qubit-string message to
many distant receivers, via the control of many agents in a network. A brief
discussion and the concluding summary are given in Sec. V.

\begin{center}
{\bf II. TELEPORTATION OF MULTI{\it -}QUBIT INFORMATION TO A DISTANT
RECEIVER VIA THE CONTROL OF ONE AGENT }
\end{center}

Suppose that Alice holds a string of message qubits labeled by $1,2,...,m$,
which is initially in the state $\prod_{i=1}^m(\alpha _i\left|
0\right\rangle _i+\beta _i\left| 1\right\rangle _i)$. She wishes to send the 
$m$-qubit information to a distant receiver (Bob) via the control of one
agent (Carol), such that Bob can get the complete information carried by
each message qubit only if Carol collaborates. This can be done by the
following procedure:

Firstly, Alice prepares the following EPR entangled state through local
logic gates 
\begin{eqnarray}
&&\ \ \ \ \ \ \ \ \ \prod_{i=1}^m(\left| 00\right\rangle _{i^{\prime
}i^{\prime \prime }}+\left| 11\right\rangle _{i^{\prime }i^{\prime \prime
}})\otimes (\left| 00\right\rangle _{ac}+\left| 11\right\rangle _{ac}) 
\nonumber \\
&&\ \ \ \ \ \ \ \ \ +\prod_{i=1}^m(\left| 00\right\rangle _{i^{\prime
}i^{\prime \prime }}-\left| 11\right\rangle _{i^{\prime }i^{\prime \prime
}})\otimes (\left| 00\right\rangle _{ac}-\left| 11\right\rangle _{ac}),
\end{eqnarray}
and then sends EPR qubit $c$ to Carol and $m$ EPR qubits ($1^{\prime \prime
},2^{\prime \prime },...,m^{\prime \prime }$) to Bob, while keeping the
other $m+1$ EPR qubits ($1^{\prime },2^{\prime },...,m^{\prime }$) and $a$
to herself. The state of the whole system is given by 
\begin{eqnarray}
&&\ \ \ \ \ \ \prod_{i=1}^m\left[ (\alpha _i\left| 0\right\rangle _i+\beta
_i\left| 1\right\rangle _i)(\left| 00\right\rangle _{i^{\prime }i^{\prime
\prime }}+\left| 11\right\rangle _{i^{\prime }i^{\prime \prime }})\right]
\otimes (\left| 00\right\rangle _{ac}+\left| 11\right\rangle _{ac}) 
\nonumber \\
&&\ \ \ \ \ \ +\prod_{i=1}^m\left[ (\alpha _i\left| 0\right\rangle _i+\beta
_i\left| 1\right\rangle _i)(\left| 00\right\rangle _{i^{\prime }i^{\prime
\prime }}-\left| 11\right\rangle _{i^{\prime }i^{\prime \prime }})\right]
\otimes (\left| 00\right\rangle _{ac}-\left| 11\right\rangle _{ac}),
\end{eqnarray}
which can be written as 
\begin{eqnarray}
&&\ \ \prod_{i=1}^m\left[ \left| \phi _{ii^{\prime }}^{+}\right\rangle
(\alpha _i\left| 0\right\rangle _{i^{\prime \prime }}+\beta _i\left|
1\right\rangle _{i^{\prime \prime }})+\left| \phi _{ii^{\prime
}}^{-}\right\rangle (\alpha _i\left| 0\right\rangle _{i^{\prime \prime
}}-\beta _i\left| 1\right\rangle _{i^{\prime \prime }})\right.  \nonumber \\
&&\ \ \left. +\left| \psi _{ii^{\prime }}^{+}\right\rangle (\alpha _i\left|
1\right\rangle _{i^{\prime \prime }}+\beta _i\left| 0\right\rangle
_{i^{\prime \prime }})+\left| \psi _{ii^{\prime }}^{-}\right\rangle (\alpha
_i\left| 1\right\rangle _{i^{\prime \prime }}-\beta _i\left| 0\right\rangle
_{i^{\prime \prime }})\right]  \nonumber \\
&&\ \ \otimes (\left| 00\right\rangle _{ac}+\left| 11\right\rangle _{ac}) 
\nonumber \\
&&\ \ +\prod_{i=1}^m\left[ \left| \phi _{ii^{\prime }}^{+}\right\rangle
(\alpha _i\left| 0\right\rangle _{i^{\prime \prime }}-\beta _i\left|
1\right\rangle _{i^{\prime \prime }})+\left| \phi _{ii^{\prime
}}^{-}\right\rangle (\alpha _i\left| 0\right\rangle _{i^{\prime \prime
}}+\beta _i\left| 1\right\rangle _{i^{\prime \prime }})\right.  \nonumber \\
&&\ \ \left. +\left| \psi _{ii^{\prime }}^{+}\right\rangle (-\alpha _i\left|
1\right\rangle _{i^{\prime \prime }}+\beta _i\left| 0\right\rangle
_{i^{\prime \prime }})+\left| \psi _{ii^{\prime }}^{-}\right\rangle (-\alpha
_i\left| 1\right\rangle _{i^{\prime \prime }}-\beta _i\left| 0\right\rangle
_{i^{\prime \prime }})\right]  \nonumber \\
&&\ \ \otimes (\left| 00\right\rangle _{ac}-\left| 11\right\rangle _{ac}).
\end{eqnarray}

Here and below, the subscripts $ii^{\prime }=11^{\prime },22^{\prime
},33^{\prime }...;$ $i^{\prime }i^{\prime \prime }=1^{\prime }1^{\prime
\prime },2^{\prime }2^{\prime \prime },3^{\prime }3^{\prime \prime }...;$
and $i^{\prime \prime }=1^{\prime \prime },2^{\prime \prime },3^{\prime
\prime }...;$ for $i=1,2,3...$. In addition, normalized factors throughout
this paper are omitted for simplicity. The states $\left| \phi _{ii^{\prime
}}^{+}\right\rangle ,\left| \phi _{ii^{\prime }}^{-}\right\rangle ,\left|
\psi _{ii^{\prime }}^{+}\right\rangle ,$ and $\left| \psi _{ii^{\prime
}}^{-}\right\rangle $ involved in (4) are the four Bell states for the qubit
pair ($i,i^{\prime }$), which form a set of the complete orthogonal basis in
the 4D Hilbert space of the two qubits $i$ and $i^{\prime },$ and take the
following expressions 
\begin{eqnarray}
\left| \phi _{ii^{\prime }}^{\pm }\right\rangle &=&\left| 00\right\rangle
_{ii^{\prime }}\pm \left| 11\right\rangle _{ii^{\prime }},  \nonumber \\
\left| \psi _{ii^{\prime }}^{\pm }\right\rangle &=&\left| 01\right\rangle
_{ii^{\prime }}\pm \left| 10\right\rangle _{ii^{\prime }},
\end{eqnarray}
respectively.

Secondly, Alice performs a series of two-qubit Bell-state measurements
respectively on qubit pairs ($1,1^{\prime }$), ($2,2^{\prime }$),...($%
m,m^{\prime }$). After that, one has

\begin{equation}
\ \ \ \left| \psi \right\rangle (\left| 00\right\rangle _{ac}+\left|
11\right\rangle _{ac})+\left| \psi ^{\prime }\right\rangle (\left|
00\right\rangle _{ac}-\left| 11\right\rangle _{ac})
\end{equation}
with 
\begin{equation}
\left| \psi \right\rangle =\prod_{i=1}^m\left| \psi \right\rangle
_{i^{\prime \prime }},\qquad \left| \psi ^{\prime }\right\rangle
=\prod_{i=1}^m\left| \psi ^{\prime }\right\rangle _{i^{\prime \prime }},
\end{equation}
where $\left| \psi \right\rangle $ and $\left| \psi ^{\prime }\right\rangle $
are the states for the $m$ qubits ($1^{\prime \prime },2^{\prime \prime
},...,m^{\prime \prime }$) belonging to Bob, while $\left| \psi
\right\rangle _{i^{\prime \prime }}$ and $\left| \psi \right\rangle
_{i^{\prime \prime }}^{\prime }$ are the states of Bob's qubit $i^{\prime
\prime }.$ From (4), one can see that the states $\left| \psi \right\rangle
_{i^{\prime \prime }}$ and $\left| \psi \right\rangle _{i^{\prime \prime
}}^{\prime }$ depend on the outcome of Alice's Bell-state measurement on the
qubit pair ($i,i^{\prime }$), and are given by 
\begin{equation}
\left| \psi \right\rangle _{i^{\prime \prime }}=\left\{ 
\begin{array}{c}
\alpha _i\left| 0\right\rangle _{i^{\prime \prime }}+\beta _i\left|
1\right\rangle _{i^{\prime \prime }}\text{ \quad for }P_{ii^{\prime
}}=\left| \phi _{ii^{\prime }}^{+}\right\rangle \left\langle \phi
_{ii^{\prime }}^{+}\right| , \\ 
\alpha _i\left| 0\right\rangle _{i^{\prime \prime }}-\beta _i\left|
1\right\rangle _{i^{\prime \prime }}\text{ \quad for }P_{ii^{\prime
}}=\left| \phi _{ii^{\prime }}^{-}\right\rangle \left\langle \phi
_{ii^{\prime }}^{-}\right| , \\ 
\alpha _i\left| 1\right\rangle _{i^{\prime \prime }}+\beta _i\left|
0\right\rangle _{i^{\prime \prime }}\text{ \quad for }P_{ii^{\prime
}}=\left| \psi _{ii^{\prime }}^{+}\right\rangle \left\langle \psi
_{ii^{\prime }}^{+}\right| , \\ 
\alpha _i\left| 1\right\rangle _{i^{\prime \prime }}-\beta _i\left|
0\right\rangle _{i^{\prime \prime }}\text{ \quad for }P_{ii^{\prime
}}=\left| \psi _{ii^{\prime }}^{-}\right\rangle \left\langle \psi
_{ii^{\prime }}^{-}\right| ;
\end{array}
\right.
\end{equation}
\begin{equation}
\left| \psi ^{\prime }\right\rangle _{i^{\prime \prime }}=\left\{ 
\begin{array}{c}
\alpha _i\left| 0\right\rangle _{i^{\prime \prime }}-\beta _i\left|
1\right\rangle _{i^{\prime \prime }}\text{ \quad for }P_{ii^{\prime
}}=\left| \phi _{ii^{\prime }}^{+}\right\rangle \left\langle \phi
_{ii^{\prime }}^{+}\right| , \\ 
\alpha _i\left| 0\right\rangle _{i^{\prime \prime }}+\beta _i\left|
1\right\rangle _{i^{\prime \prime }}\text{ \quad for }P_{ii^{\prime
}}=\left| \phi _{ii^{\prime }}^{-}\right\rangle \left\langle \phi
_{ii^{\prime }}^{-}\right| , \\ 
-\alpha _i\left| 1\right\rangle _{i^{\prime \prime }}+\beta _i\left|
0\right\rangle _{i^{\prime \prime }}\text{ \quad for }P_{ii^{\prime
}}=\left| \psi _{ii^{\prime }}^{+}\right\rangle \left\langle \psi
_{ii^{\prime }}^{+}\right| , \\ 
-\alpha _i\left| 1\right\rangle _{i^{\prime \prime }}-\beta _i\left|
0\right\rangle _{i^{\prime \prime }}\text{ \quad for }P_{ii^{\prime
}}=\left| \psi _{ii^{\prime }}^{-}\right\rangle \left\langle \psi
_{ii^{\prime }}^{-}\right| .
\end{array}
\right.
\end{equation}
where $P_{ii^{\prime }}$ is a projector onto the Bell state $\left| \phi
_{ii^{\prime }}^{+}\right\rangle ,\left| \phi _{ii^{\prime
}}^{-}\right\rangle ,\left| \psi _{ii^{\prime }}^{+}\right\rangle ,$ or $%
\left| \psi _{ii^{\prime }}^{-}\right\rangle $ of the two qubits in the pair
($i,i^{\prime }$).

The results (8) and (9) show that according to the outcome of Alice's
Bell-state measurement on the pair ($i,i^{\prime }$), Bob can always recover
the original state $\alpha _i\left| 0\right\rangle _i+\beta _i\left|
1\right\rangle _i$ of the message qubit $i$ from the state $\left| \psi
\right\rangle _{i^{\prime \prime }}$ or $\left| \psi ^{\prime }\right\rangle
_{i^{\prime \prime }}$ of his qubit $i^{\prime \prime }$, by performing a
single-qubit operation on the qubit $i^{\prime \prime }.$ For instance, Bob
obtains $U\left| \psi \right\rangle _{i^{\prime \prime }}\rightarrow \alpha
_i\left| 0\right\rangle _{i^{\prime \prime }}+\beta _i\left| 1\right\rangle
_{i^{\prime \prime }}$ and $U^{\prime }\left| \psi ^{\prime }\right\rangle
_{i^{\prime \prime }}\rightarrow \alpha _i\left| 0\right\rangle _{i^{\prime
\prime }}+\beta _i\left| 1\right\rangle _{i^{\prime \prime }}$, where $U$
and $U^{\prime }$ are unitary operators and $U,U^{\prime }=I,\sigma
_z;\sigma _z,I;$ $\sigma _x,\sigma _y;$ or $\sigma _y,\sigma _x,$ for the
case of Alice measuring the pair ($i,i^{\prime }$) in the Bell state $\left|
\phi _{ii^{\prime }}^{+}\right\rangle ,\left| \phi _{ii^{\prime
}}^{-}\right\rangle ,\left| \psi _{ii^{\prime }}^{+}\right\rangle ,$ or $%
\left| \psi _{ii^{\prime }}^{-}\right\rangle $. Here, $\sigma _x,\sigma _y,$
and $\sigma _z$ are Pauli operators which correspond to the rotations by $%
\pi $ rad about the $x$, $y$, and $z$ axes, respectively, and $I$ is an
identity operator.

Thirdly, Alice and Carol perform a Hadamard transformation on their
respective qubits $a$ and $c.$ As a result, the state $\left|
00\right\rangle _{ac}-\left| 11\right\rangle _{ac}$ goes to $\left|
01\right\rangle _{ac}+\left| 10\right\rangle _{ac}$ while the state $\left|
00\right\rangle _{ac}+\left| 11\right\rangle _{ac}$ remains unchanged. Thus,
the state (6) will, after Alice's and Carol's Hadamard transformations,
change into 
\begin{equation}
\ \ \left| \psi \right\rangle (\left| 00\right\rangle _{ac}+\left|
11\right\rangle _{ac})+\left| \psi ^{\prime }\right\rangle (\left|
01\right\rangle _{ac}+\left| 10\right\rangle _{ac}).
\end{equation}

Fourthly, Alice and Carol make a measurement on their respective qubits $a$
and $c,$ and then each sends the measurement result (one-bit classical
message) to Bob. One can see from (10) that if Bob knows that Alice and
Carol both measured their qubits in the state $\left| 0\right\rangle $ or $%
\left| 1\right\rangle ,$ he can predict that his $m$ qubits ($1^{\prime
\prime },2^{\prime \prime },...,m^{\prime \prime }$) must be in the state $%
\left| \psi \right\rangle $. On the other hand, expression (10) shows that
if Bob knows that Alice measured her qubit in the state $\left|
0\right\rangle $ ($\left| 1\right\rangle $) while Carol measured his qubit
in the state $\left| 1\right\rangle $ ($\left| 0\right\rangle $), he knows
that his $m$ qubits must be in the state $\left| \psi ^{\prime
}\right\rangle $. Therefore, according to the measurement results from Alice
and Carol, Bob can predict whether his $m$ qubits are in $\left| \psi
\right\rangle $ or $\left| \psi ^{\prime }\right\rangle $.

Lastly, note that the state $\left| \psi \right\rangle $($\left| \psi
^{\prime }\right\rangle $), described in (7), is a product of
individual-qubit states $\left| \psi \right\rangle _{1^{\prime \prime
}},\left| \psi \right\rangle _{2^{\prime \prime }},...,\left| \psi
\right\rangle _{m^{\prime \prime }}$ ($\left| \psi ^{\prime }\right\rangle
_{1^{\prime \prime }},\left| \psi ^{\prime }\right\rangle _{2^{\prime \prime
}},...,\left| \psi ^{\prime }\right\rangle _{m^{\prime \prime }}$) for the
qubits ($1^{\prime \prime },2^{\prime \prime },...,m^{\prime \prime }$).
And, as addressed above, Bob can recover the original state (i.e., $\alpha
_i\left| 0\right\rangle _i+\beta _i\left| 1\right\rangle _i$ of message
qubit $i$) from the state $\left| \psi \right\rangle _{i^{\prime \prime }}$
or $\left| \psi ^{\prime }\right\rangle _{i^{\prime \prime }}$ of his qubit $%
i^{\prime \prime },$ based on the outcome of Alice's Bell-state measurement
on the qubit pair ($i,i^{\prime }$) and via a single-qubit unitary
transformation $U$ or $U^{\prime }$ on the qubit $i^{\prime \prime }$.
Hence, Bob can always reconstruct the original state of $m$ message qubits ($%
1,2,...,m$) from the state $\left| \psi \right\rangle $ or $\left| \psi
^{\prime }\right\rangle $ of his $m$ qubits ($1^{\prime \prime },2^{\prime
\prime },...,m^{\prime \prime }$), according to the outcome of Alice's
Bell-state measurements on the qubit pairs ($1,1^{\prime }$), ($2,2^{\prime
} $),..., ($m,m^{\prime }$) and through his local single-qubit operations.

In what follows, our purpose is to show that when Carol does not
collaborate, it is impossible for Bob to gain the full quantum message.
Examining the state (6), we see that when only Alice performs a Hadamard
transformation on her qubit $a,$ the state (6) will change into 
\begin{eqnarray}
&&\ \ \left[ (\left| \psi \right\rangle +\left| \psi ^{\prime }\right\rangle
)\left| 0\right\rangle _c+(\left| \psi \right\rangle -\left| \psi ^{\prime
}\right\rangle )\left| 1\right\rangle _c\right] \left| 0\right\rangle _a 
\nonumber \\
&&\ +\ \left[ (\left| \psi \right\rangle +\left| \psi ^{\prime
}\right\rangle )\left| 0\right\rangle _c-(\left| \psi \right\rangle -\left|
\psi ^{\prime }\right\rangle )\left| 1\right\rangle _c\right] \left|
1\right\rangle _a,
\end{eqnarray}
which implies that whether Alice measures her qubit $a$ in the state $\left|
0\right\rangle $ or $\left| 1\right\rangle $, the density operator of the $m$
qubits ($1^{\prime \prime },2^{\prime \prime },...,m^{\prime \prime }$)
belonging to Bob will, after tracing over Carol's qubit $c$, be given by 
\begin{equation}
\rho =(\left| \psi \right\rangle +\left| \psi ^{\prime }\right\rangle
)(\left\langle \psi \right| +\left\langle \psi ^{\prime }\right| )+(\left|
\psi \right\rangle -\left| \psi ^{\prime }\right\rangle )(\left\langle \psi
\right| -\left\langle \psi ^{\prime }\right| ).
\end{equation}
The result (12) shows that the $m$ qubits ($1^{\prime \prime },2^{\prime
\prime },...,m^{\prime \prime }$) are in a mixed state, in which they are in
a superposition state $\left| \psi \right\rangle +\left| \psi ^{\prime
}\right\rangle $ with a probability $p_1=\frac{\left| \left| \psi
\right\rangle +\left| \psi ^{\prime }\right\rangle \right| ^2}{2\left(
\left\langle \psi \right| \left. \psi \right\rangle +\left\langle \psi
^{\prime }\right| \left. \psi ^{\prime }\right\rangle \right) }$ while being
in the other superposition state $\left| \psi \right\rangle -\left| \psi
^{\prime }\right\rangle $ with a probability $p_2=\frac{\left| \left| \psi
\right\rangle -\left| \psi ^{\prime }\right\rangle \right| ^2}{2\left(
\left\langle \psi \right| \left. \psi \right\rangle +\left\langle \psi
^{\prime }\right| \left. \psi ^{\prime }\right\rangle \right) }.$

Based on Eqs. (7)-(9), one can express the states $\left| \psi \right\rangle
+\left| \psi ^{\prime }\right\rangle $ and $\left| \psi \right\rangle
-\left| \psi ^{\prime }\right\rangle $ involved in Eq. (12) as follows 
\begin{eqnarray}
\left| \psi \right\rangle \pm \left| \psi ^{\prime }\right\rangle &=&\left| 
\widetilde{\psi }\right\rangle (\alpha _t\left| 0\right\rangle _{t^{\prime
\prime }}+\beta _t\left| 1\right\rangle _{t^{\prime \prime }})\pm \left| 
\widetilde{\psi }^{\prime }\right\rangle (\alpha _t\left| 0\right\rangle
_{t^{\prime \prime }}-\beta _t\left| 1\right\rangle _{t^{\prime \prime }})%
\text{ for }P_{tt^{\prime }}=\left| \phi _{tt^{\prime }}^{+}\right\rangle
\left\langle \phi _{tt^{\prime }}^{+}\right| ,  \nonumber \\
\left| \psi \right\rangle \pm \left| \psi ^{\prime }\right\rangle &=&\left| 
\widetilde{\psi }\right\rangle (\alpha _t\left| 0\right\rangle _{t^{\prime
\prime }}-\beta _t\left| 1\right\rangle _{t^{\prime \prime }})\pm \left| 
\widetilde{\psi }^{\prime }\right\rangle (\alpha _t\left| 0\right\rangle
_{t^{\prime \prime }}+\beta _t\left| 1\right\rangle _{t^{\prime \prime }})%
\text{ for }P_{tt^{\prime }}=\left| \phi _{tt^{\prime }}^{-}\right\rangle
\left\langle \phi _{tt^{\prime }}^{-}\right| ,  \nonumber \\
\left| \psi \right\rangle \pm \left| \psi ^{\prime }\right\rangle &=&\left| 
\widetilde{\psi }\right\rangle (\alpha _t\left| 1\right\rangle _{t^{\prime
\prime }}+\beta _t\left| 0\right\rangle _{t^{\prime \prime }})\pm \left| 
\widetilde{\psi }^{\prime }\right\rangle (-\alpha _t\left| 1\right\rangle
_{t^{\prime \prime }}+\beta _t\left| 0\right\rangle _{t^{\prime \prime }})%
\text{ for }P_{tt^{\prime }}=\left| \psi _{tt^{\prime }}^{+}\right\rangle
\left\langle \psi _{tt^{\prime }}^{+}\right| ,  \nonumber \\
\left| \psi \right\rangle \pm \left| \psi ^{\prime }\right\rangle &=&\left| 
\widetilde{\psi }\right\rangle (\alpha _t\left| 1\right\rangle _{t^{\prime
\prime }}-\beta _t\left| 0\right\rangle _{t^{\prime \prime }})\pm \left| 
\widetilde{\psi }^{\prime }\right\rangle (-\alpha _t\left| 1\right\rangle
_{t^{\prime \prime }}-\beta _t\left| 0\right\rangle _{t^{\prime \prime }})%
\text{ for }P_{tt^{\prime }}=\left| \psi _{tt^{\prime }}^{-}\right\rangle
\left\langle \psi _{tt^{\prime }}^{-}\right| ,
\end{eqnarray}
where the subscript $t^{\prime \prime }$ represents any one of the $m$
qubits ($1^{\prime \prime },2^{\prime \prime },...,m^{\prime \prime }$)
belonging to Bob, and the subscripts $t$ and $t^{\prime }$ denote Alice's
message qubit and her EPR qubit (corresponding to Bob's qubit $t^{\prime
\prime }$), respectively. In Eq. (13), we further note that $\left| 
\widetilde{\psi }\right\rangle =\prod_k\left| \psi \right\rangle _{k^{\prime
\prime }}$ and $\left| \widetilde{\psi }^{\prime }\right\rangle
=\prod_k\left| \psi ^{\prime }\right\rangle _{k^{\prime \prime }}$ ($k\neq t$%
) are the states of the remaining $m-1$ qubits belonging to Bob (after
excluding the qubit $t^{\prime \prime }$). Here, $\left| \psi \right\rangle
_{k^{\prime \prime }}$ and $\left| \psi ^{\prime }\right\rangle _{k^{\prime
\prime }}$ are the states of qubit $k^{\prime \prime }$ ($k^{\prime \prime
}\neq t^{\prime \prime }$), which depend on the outcome of Alice's
Bell-state measurement on the pair ($k,k^{\prime }$) and take the form of
(8) and (9), respectively. From Eqs. (12) and (13), it is easily shown that
for each outcome ($\left| \phi _{tt^{\prime }}^{+}\right\rangle ,\left| \phi
_{tt^{\prime }}^{-}\right\rangle ,\left| \psi _{tt^{\prime
}}^{+}\right\rangle ,$ or $\left| \psi _{tt^{\prime }}^{-}\right\rangle $)
of Alice's Bell-state measurement on the pair ($t,t^{\prime }$), the density
operator for the remaining $m-1$ qubits belonging to Bob is, after tracing
over the qubit $t^{\prime \prime },$ given by 
\begin{eqnarray}
\widetilde{\rho } &=&tr_{t^{\prime \prime }}(\rho )  \nonumber \\
\ &=&\left( \left| \widetilde{\psi }\right\rangle +\left| \widetilde{\psi }%
^{\prime }\right\rangle \right) \left( \left\langle \widetilde{\psi }\right|
+\left\langle \widetilde{\psi }^{\prime }\right| \right) \left| \alpha
_t\right| ^2+\left( \left| \widetilde{\psi }\right\rangle +\left| \widetilde{%
\psi }^{\prime }\right\rangle \right) \left( \left\langle \widetilde{\psi }%
\right| +\left\langle \widetilde{\psi }^{\prime }\right| \right) \left|
\beta _t\right| ^2  \nonumber \\
&&\ \ \ \ \ +\left( \left| \widetilde{\psi }\right\rangle -\left| \widetilde{%
\psi }^{\prime }\right\rangle \right) \left( \left\langle \widetilde{\psi }%
\right| -\left\langle \widetilde{\psi }^{\prime }\right| \right) \left|
\alpha _t\right| ^2+\left( \left| \widetilde{\psi }\right\rangle -\left| 
\widetilde{\psi }^{\prime }\right\rangle \right) \left( \left\langle 
\widetilde{\psi }\right| -\left\langle \widetilde{\psi }^{\prime }\right|
\right) \left| \beta _t\right| ^2  \nonumber \\
\ &=&\left( \left| \widetilde{\psi }\right\rangle +\left| \widetilde{\psi }%
^{\prime }\right\rangle \right) \left( \left\langle \widetilde{\psi }\right|
+\left\langle \widetilde{\psi }^{\prime }\right| \right) +\left( \left| 
\widetilde{\psi }\right\rangle -\left| \widetilde{\psi }^{\prime
}\right\rangle \right) \left( \left\langle \widetilde{\psi }\right|
-\left\langle \widetilde{\psi }^{\prime }\right| \right) ,
\end{eqnarray}
where we have used $\left| \alpha _t\right| ^2+$ $\left| \beta _t\right|
^2=1.$ Eq. (14) implies that the density operator, for the remaining $m-1$
``non-traced'' qubits belonging to Bob, has the same form as (12).
Therefore, repeating the above single-qubit tracing procedure, one finds
that the density operator for any qubit $i^{\prime \prime }$ belonging to
Bob ($i^{\prime \prime }=1^{\prime \prime },2^{\prime \prime },...,$ or $%
m^{\prime \prime }$) can, after tracing over Bob's other $m-1$ qubits, be
written as 
\begin{equation}
\rho _{i^{\prime \prime }}=\left| \alpha _i\right| ^2\left| 0\right\rangle
\left\langle 0\right| +\left| \beta _i\right| ^2\left| 1\right\rangle
\left\langle 1\right|
\end{equation}
in the case when Alice measures the qubit pair ($i,$ $i^{\prime }$) in
either Bell state $\left| \phi _{ii^{\prime }}^{+}\right\rangle $ or $\left|
\phi _{ii^{\prime }}^{-}\right\rangle .$ On the other hand, 
\begin{equation}
\rho _{i^{\prime \prime }}=\left| \alpha _i\right| ^2\left| 1\right\rangle
\left\langle 1\right| +\left| \beta _i\right| ^2\left| 0\right\rangle
\left\langle 0\right|
\end{equation}
in the case when the pair ($i,$ $i^{\prime }$) is measured in the Bell state 
$\left| \psi _{ii^{\prime }}^{+}\right\rangle $ or $\left| \psi _{ii^{\prime
}}^{-}\right\rangle .$ The above process demonstrates that the density
operator (15) or (16) depends only on the outcome of Alice's Bell-state
measurement on the pair ($i,$ $i^{\prime }$), but independent of the outcome
of Alice's Bell-state measurement on all other pairs.

Eqs. (15) and (16) demonstrate the following results. First, any qubit $%
i^{\prime \prime }$ belonging to Bob is in a mixed state, in which it is in
the state $\left| 0\right\rangle $ with a probability $\left| \alpha
_i\right| ^2$ or $\left| \beta _i\right| ^2$ (depending on Alice's
Bell-state measurement outcome), while being in the state $\left|
1\right\rangle $ with a probability $\left| \beta _i\right| ^2$ or $\left|
\alpha _i\right| ^2.$ Second, Bob has amplitude information about Alice's
each message qubit, but knows nothing about its phase. Therefore, in general
(i.e., $\alpha _i\neq 0$ or $\beta _i\neq 0$ for $i=1,2,...,m$), Bob can not
fully restore the original state of each message qubit belonging to Alice,
even if one agent does not cooperate with him.

It is necessary for us to compare the present method with the method in
[13,14]. From the description above, we conclude that to teleport $m$-qubit
information to a distant receiver via the control of one agent, the present
method requires only:

(i) $2(m+1)$ auxiliary qubits for the preparation of the state (2);

(ii) one qubit being assigned to the controller;

(iii) one single-qubit Hadamard transformation and one single-qubit
measurement being performed by the controller; and

(iv) one-bit classical message being sent to the receiver.

However, as mentioned in the introduction, to implement the present task,
the method in [13,14] will require:

(i) $3m$ auxiliary qubits for preparing $m$ copies of a three-qubit GHZ
state;

(ii) $m$ qubits being assigned to the controller;

(iii) $m$ single-qubit Hadamard transformations and $m$ single-qubit
measurements being performed by the controller; and

(iv) $m$-bit classical messages being sent to the receiver by the controller.

The above analysis demonstrates that for the case of $m=1$, the present
protocol is trivial, since it requires 4 auxiliary qubits while the method
in [13,14] only needs 3 auxiliary qubits. However, the advantage for the
present proposal appears when $m=2,$ because it requires the same number of
auxiliary qubits but less local operation and classical communication,
compared with the method in [13,14]. Moreover, when $m=3,$ the number of
auxiliary qubits required in the present method becomes smaller than that
using the method in [13,14]. One can clearly see that the advantage of the
present method becomes apparent with the increment of $m.$ Especially, when $%
m$ is a large number, the required auxiliary qubit resources, local
operations by the controller, and classical communication between the
controller and the receiver are greatly reduced in the present approach.

On a final note, we point out that the number of Alice's Bell-state
measurements needed in the present protocol is the same as that required by
the method in [13,14]. This is obvious, since using the method in [13,14],
Alice also needs to perform a series of Bell-state measurements, each acting
on one message qubit and one GHZ qubit.

\begin{center}
{\bf III. TELEPORTATION OF MULTI-QUBIT INFORMATION TO A DISTANT RECEIVER VIA
THE CONTROL OF MANY AGENTS}
\end{center}

In this section, we discuss how to decompose multi-qubit GHZ states and how
to generalize the above method to the teleportation of multi-qubit
information via the control of many agents.

\begin{center}
{\bf A. Decomposition of multi-qubit GHZ states}
\end{center}

In the past few years, GHZ states have been extensively studied by many
researchers. They play an important role in quantum information processing
and communication. Many theoretical proposals have appeared for the
generation of multi-qubit GHZ states. Moreover, it has been reported that up
to four-qubit GHZ states were experimentally prepared with polarized-state
photons [23] and trapped ions [24]. As especially relevant to this work, we
consider the following two types of ($n+1$)-qubit GHZ states: 
\begin{eqnarray}
\left| GHZ\right\rangle _{+} &=&\left| 00...0\right\rangle +\left|
11...1\right\rangle , \\
\left| GHZ\right\rangle _{-} &=&\left| 00...0\right\rangle -\left|
11...1\right\rangle .
\end{eqnarray}
We find that if a Hadamard transformation is performed on each qubit, the
states (17) and (18) will be decomposed, respectively, into 
\begin{equation}
\left| GHZ\right\rangle _{+}\rightarrow \sum_{\{x_l\}}\left|
\{x_l\}\right\rangle \left| 0\right\rangle +\sum_{\{y_l\}}\left|
\{y_l\}\right\rangle \left| 1\right\rangle ,
\end{equation}
\begin{equation}
\left| GHZ\right\rangle _{-}\rightarrow \sum_{\{x_l\}}\left|
\{x_l\}\right\rangle \left| 1\right\rangle +\sum_{\{y_l\}}\left|
\{y_l\}\right\rangle \left| 0\right\rangle ,
\end{equation}
where $\left| \{x_l\}\right\rangle =\left| x_1x_2...x_n\right\rangle $ and $%
\left| \left\{ y_l\right\} \right\rangle =\left| y_1y_2...y_n\right\rangle $
are computational basis states of the first $n$ qubits ($x_l,y_l\in
\{0,1\};l=1,2,...n$), and $\sum_{\{x_l\}}\left| \{x_l\}\right\rangle $ ($%
\sum_{\{y_l\}}\left| \{y_l\}\right\rangle $) is a sum over all possible
basis states $\left| \{x_l\}\right\rangle $ ($\left| \{y_l\}\right\rangle $)
each containing an $even$ ($odd$) number of ``1''s. For instance, when $n=4,$
$\sum_{\{x_l\}}\left| \{x_l\}\right\rangle =\left| 0000\right\rangle +\left|
1100\right\rangle +\left| 1010\right\rangle +\cdot \cdot \cdot +\left|
1111\right\rangle .$ Note that the number of the basis states $\left|
\{x_l\}\right\rangle $ is the same as that of the basis states $\left|
\{y_l\}\right\rangle $, thus the states (19) and\ (20) on the right side
both have the same normalized factors.

\begin{center}
{\bf B. Teleportation of }$m${\bf -qubit information via the control of }$n$%
{\bf \ agents}
\end{center}

Now, suppose that Alice has a string of message qubits labeled by $1,2,...,m$%
, which is initially in the state $\prod_{i=1}^m(\alpha _i\left|
0\right\rangle _i+\beta _i\left| 1\right\rangle _i)$. She wishes to send the 
$m$-qubit information to Bob via the control of $n$ agents ($A_1,A_2,...,A_n$%
) in a network, such that Bob can get the complete information of each
message qubit only if all the agents collaborate. This can be done by the
following procedure:

Firstly, Alice prepares the following EPR-GHZ entangled state through local
logic gates 
\begin{eqnarray}
&&\ \ \ \ \ \ \ \prod_{i=1}^m(\left| 00\right\rangle _{i^{\prime }i^{\prime
\prime }}+\left| 11\right\rangle _{i^{\prime }i^{\prime \prime }})\otimes
\left| GHZ\right\rangle _{+}  \nonumber \\
&&\ \ \ \ \ \ \ +\prod_{i=1}^m(\left| 00\right\rangle _{i^{\prime }i^{\prime
\prime }}-\left| 11\right\rangle _{i^{\prime }i^{\prime \prime }})\otimes
\left| GHZ\right\rangle _{-},
\end{eqnarray}
(where $\left| GHZ\right\rangle _{\pm }=\left| 00...0\right\rangle \pm
\left| 11...1\right\rangle $ are ($n+1$)-qubit GHZ states), and then she
sends the first $n$ GHZ qubits to the $n$ agents and the $m$ EPR qubits ($%
1^{\prime \prime },2^{\prime \prime },...,m^{\prime \prime }$) to Bob, while
keeping the last GHZ qubit and the other $m$ EPR qubits ($1^{\prime
},2^{\prime },...,m^{\prime }$) to herself. The state of the whole system is
given by 
\begin{eqnarray}
&&\ \ \ \ \prod_{i=1}^m\left[ (\alpha _i\left| 0\right\rangle _i+\beta
_i\left| 1\right\rangle _i)(\left| 00\right\rangle _{i^{\prime }i^{\prime
\prime }}+\left| 11\right\rangle _{i^{\prime }i^{\prime \prime }})\right]
\otimes \left| GHZ\right\rangle _{+}  \nonumber \\
&&\ \ \ \ +\prod_{i=1}^m\left[ (\alpha _i\left| 0\right\rangle _i+\beta
_i\left| 1\right\rangle _i)(\left| 00\right\rangle _{i^{\prime }i^{\prime
\prime }}-\left| 11\right\rangle _{i^{\prime }i^{\prime \prime }})\right]
\otimes \left| GHZ\right\rangle _{-}.
\end{eqnarray}

Secondly, Alice performs a series of two-qubit Bell-state measurements
respectively on $m$ qubit pairs ($1,1^{\prime }$), ($2,2^{\prime }$),...($%
m,m^{\prime }$). After that, we have

\begin{equation}
\ \ \ \left| \psi \right\rangle \left| GHZ\right\rangle _{+}+\left| \psi
^{\prime }\right\rangle \left| GHZ\right\rangle _{-},
\end{equation}
where $\left| \psi \right\rangle $ and $\left| \psi ^{\prime }\right\rangle $
are the states for the $m$ qubits ($1^{\prime \prime },2^{\prime \prime
},...,m^{\prime \prime }$) belonging to Bob. Note that the left part of the
first (second) product term in (22) is the same as that of the first
(second) product term in (3). Thus, the two states $\left| \psi
\right\rangle $ and $\left| \psi ^{\prime }\right\rangle $ here take the
same form as $\left| \psi \right\rangle $ and $\left| \psi ^{\prime
}\right\rangle $ described by (7), respectively.

Thirdly, each agent and Alice perform a Hadamard transformation on their
respective GHZ qubits. After that, based on (19) and (20), one gets from
(23) 
\begin{eqnarray}
&&\ \ \ \left| \psi \right\rangle \left[ \sum_{\left\{ x_l\right\} }\left|
\left\{ x_l\right\} \right\rangle \left| 0\right\rangle +\sum_{\left\{
x_l\right\} }\left| \left\{ y_l\right\} \right\rangle \left| 1\right\rangle
\right]  \nonumber \\
&&\ \ \ +\left| \psi ^{\prime }\right\rangle \left[ \sum_{\left\{
x_l\right\} }\left| \left\{ x_l\right\} \right\rangle \left| 1\right\rangle
+\sum_{\left\{ y_l\right\} }\left| \left\{ y_l\right\} \right\rangle \left|
0\right\rangle \right] .
\end{eqnarray}

Lastly, each agent and Alice make a measurement on their respective GHZ
qubits, and then send their measurement results to Bob. Recall the notation
of $\left| \left\{ x_l\right\} \right\rangle $ and $\left| \left\{
y_l\right\} \right\rangle $ described above, i.e., each basis state $\left|
\left\{ x_l\right\} \right\rangle $ ($\left| \left\{ y_l\right\}
\right\rangle $) contains an $even$ ($odd$) number of ``1''s. Therefore, one
sees from (24) that Bob can predict that the $m$ qubits ($1^{\prime \prime
},2^{\prime \prime },...,m^{\prime \prime }$) belonging to him must be in
the state $\left| \psi \right\rangle $ ($\left| \psi ^{\prime }\right\rangle 
$), if he knows that the outcome of the $n$ agents' measurement on their $n$
GHZ qubits contains an $even$ number of ``1''s and that Alice measured her
GHZ qubit in the state $\left| 0\right\rangle $ ($\left| 1\right\rangle $).
On the other hand, the result (24) shows that Bob knows that his $m$ qubits (%
$1^{\prime \prime },2^{\prime \prime },...,m^{\prime \prime }$) must be in
the state $\left| \psi \right\rangle $ ($\left| \psi ^{\prime }\right\rangle 
$), if he knows that the outcome of the $n$ agents' measurement includes an $%
odd$ number of ``1''s and that Alice measured her GHZ qubit in the state $%
\left| 1\right\rangle $ ($\left| 0\right\rangle $). Hence, according to the
measurement outcomes from the $n$ agents and Alice, Bob can predict whether
his $m$ qubits are in $\left| \psi \right\rangle $ or $\left| \psi ^{\prime
}\right\rangle $. As addressed in the previous section, Bob can restore the
original state of $m$ message qubits ($1,2,...,m$) from the state $\left|
\psi \right\rangle $ or $\left| \psi ^{\prime }\right\rangle $ of the $m$
qubits ($1^{\prime \prime },2^{\prime \prime },...,m^{\prime \prime }$),
according to the Bell-state measurement outcome from Alice and through his
local single-qubit logic operations.

We have shown, based on (23), that the quantum message originally carried by
the $m$ message qubits ($1,2,...,m$) can be recovered by Bob, as long as
each agent performs a Hadamard transformation and then a measurement on
his/her qubit. Now let us focus on the problem that Bob can not gain the
full quantum message even if one agent does not collaborate. To see this,
let us go back to the state (23). This state can be rewritten as 
\begin{eqnarray}
&&\ \ \ \ \left| \psi \right\rangle \left[ \left( \left| \phi
^{+}\right\rangle +\left| \phi ^{-}\right\rangle \right) \left|
0\right\rangle _{A_i}+\left( \left| \phi ^{+}\right\rangle -\left| \phi
^{-}\right\rangle \right) \left| 1\right\rangle _{A_i}\right]  \nonumber \\
&&\ \ \ \ +\left| \psi ^{\prime }\right\rangle \left[ \left( \left| \phi
^{+}\right\rangle +\left| \phi ^{-}\right\rangle \right) \left|
0\right\rangle _{A_i}-\left( \left| \phi ^{+}\right\rangle -\left| \phi
^{-}\right\rangle \right) \left| 1\right\rangle _{A_i}\right] ,
\end{eqnarray}
where $\left| 0\right\rangle _{A_i}$ and $\left| 1\right\rangle _{A_i}$ are
the two logic states of the GHZ qubit belonging to agent $A_i$ ($i=1,2,...,$
or $n$), while $\left| \phi ^{+}\right\rangle $ and $\left| \phi
^{-}\right\rangle ,$ taking the form of (17) and (18) respectively, are the
GHZ states of the remaining $n$ GHZ qubits belonging to other $n-1$ agents
and Alice.

Assume the agent $A_i$ does not collaborate with Bob. When the other $n-1$
agents and Alice perform a Hadamard transformation on their respective GHZ
qubits, it follows from (19) and (20) that the states $\left| \phi
^{+}\right\rangle $ and $\left| \phi ^{-}\right\rangle $ will be transformed
into 
\begin{eqnarray}
\left| \phi ^{+}\right\rangle &\rightarrow &\sum_{\left\{ x_l^{\prime
}\right\} }\left| \left\{ x_l^{\prime }\right\} \right\rangle \left|
0\right\rangle _a+\sum_{\left\{ y_l^{\prime }\right\} }\left| \left\{
y_l^{\prime }\right\} \right\rangle \left| 1\right\rangle _a,  \nonumber \\
\left| \phi ^{-}\right\rangle &\rightarrow &\sum_{\left\{ x_l^{\prime
}\right\} }\left| \left\{ x_l^{\prime }\right\} \right\rangle \left|
1\right\rangle _a+\sum_{\left\{ y_l^{\prime }\right\} }\left| \left\{
y_l^{\prime }\right\} \right\rangle \left| 0\right\rangle _a,
\end{eqnarray}
where the subscript $a$ represents the GHZ qubit belonging to Alice; $\left|
\{x_l^{\prime }\}\right\rangle =\left| x_1^{\prime }x_2^{\prime
}...x_{n-1}^{\prime }\right\rangle $ and $\left| \left\{ y_l^{\prime
}\right\} \right\rangle =\left| y_1^{\prime }y_2^{\prime }...y_{n-1}^{\prime
}\right\rangle $ are computational basis states of the $n-1$ GHZ qubits
belonging to the other $n-1$ agents ($x_l^{\prime },y_l^{\prime }\in
\{0,1\};l=1,2,...n-1$). Further, $\sum_{\{x_l^{\prime }\}}\left|
\{x_l^{\prime }\}\right\rangle $ ($\sum_{\{y_l^{\prime }\}}\left|
\{y_l^{\prime }\}\right\rangle $) represents a sum over all possible basis
states $\left| \{x_l^{\prime }\}\right\rangle $ ($\left| \{y_l^{\prime
}\}\right\rangle $) each containing an $even$ ($odd$) number of ``1''s. The
state (25) will, after replacing $\left| \phi ^{\pm }\right\rangle $ by
(26), change into

\begin{eqnarray}
&&\ \ \left[ (\left| \psi \right\rangle +\left| \psi ^{\prime }\right\rangle
)\left| 0\right\rangle _{A_i}+(\left| \psi \right\rangle -\left| \psi
^{\prime }\right\rangle )\left| 1\right\rangle _{A_i}\right] \sum_{\left\{
x_l^{\prime }\right\} }\left| \left\{ x_l^{\prime }\right\} \right\rangle
\left| 0\right\rangle _a  \nonumber \\
&&\ \ +\left[ (\left| \psi \right\rangle +\left| \psi ^{\prime
}\right\rangle )\left| 0\right\rangle _{A_i}-(\left| \psi \right\rangle
-\left| \psi ^{\prime }\right\rangle )\left| 1\right\rangle _{A_i}\right]
\sum_{\left\{ x_l^{\prime }\right\} }\left| \left\{ x_l^{\prime }\right\}
\right\rangle \left| 1\right\rangle _a  \nonumber \\
&&\ \ +\left[ (\left| \psi \right\rangle +\left| \psi ^{\prime
}\right\rangle )\left| 0\right\rangle _{A_i}-(\left| \psi \right\rangle
-\left| \psi ^{\prime }\right\rangle )\left| 1\right\rangle _{A_i}\right]
\sum_{\left\{ y_l^{\prime }\right\} }\left| \left\{ y_l^{\prime }\right\}
\right\rangle \left| 0\right\rangle _a  \nonumber \\
&&\ \ +\left[ (\left| \psi \right\rangle +\left| \psi ^{\prime
}\right\rangle )\left| 0\right\rangle _{A_i}+(\left| \psi \right\rangle
-\left| \psi ^{\prime }\right\rangle )\left| 1\right\rangle _{A_i}\right]
\sum_{\left\{ y_l^{\prime }\right\} }\left| \left\{ y_l^{\prime }\right\}
\right\rangle \left| 1\right\rangle _a,
\end{eqnarray}
which implies that if the other $n-1$ agents and Alice perform a measurement
on their respective GHZ qubits, the $m$ qubits ($1^{\prime \prime
},2^{\prime \prime },...,m^{\prime \prime }$) belonging to Bob will be
entangled with agent $A_i$'s GHZ qubit.

From (27), it is easily seen that for every outcome $\left| \left\{
x_l^{\prime }\right\} \right\rangle \left| 0\right\rangle _a,\left| \left\{
x_l^{\prime }\right\} \right\rangle \left| 1\right\rangle _a,\left| \left\{
y_l^{\prime }\right\} \right\rangle \left| 0\right\rangle _a,$ or $\left|
\left\{ y_l^{\prime }\right\} \right\rangle \left| 1\right\rangle _a$ of the
other $n-1$ agents' and Alice's measurements on their GHZ qubits, the
density operator of the $m$ qubits ($1^{\prime \prime },2^{\prime \prime
},...,m^{\prime \prime }$) belonging to Bob is, after tracing over agent $%
A_i $'s GHZ qubit, given by 
\begin{equation}
\rho =(\left| \psi \right\rangle +\left| \psi ^{\prime }\right\rangle
)(\left\langle \psi \right| +\left\langle \psi ^{\prime }\right| )+(\left|
\psi \right\rangle -\left| \psi ^{\prime }\right\rangle )(\left\langle \psi
\right| -\left\langle \psi ^{\prime }\right| ).
\end{equation}
Since Eq. (28) takes the same form as (12), one can obtain the same results
(13), (14), (15), and (16) as described above. Therefore, Bob can fully
restore the original state of $m$ message qubits ($1,2,...,m$), only if all
the agents collaborate with him.

In summary, to teleport $m$-qubit information to a distant receiver via the
control of $n$ agents, the present method requires only:

(i) $2m+n+1$ auxiliary qubits for preparing the state (21);

(ii) one qubit being distributed to each agent;

(iii) one single-qubit Hadamard transformation and one single-qubit
measurement being performed by each agent; and

(iv) one-bit classical message being sent to the receiver by each agent.

In contrast, to implement the same task, the method in [13,14] requires:

(i) $m(n+2)$ auxiliary qubits for preparing $m$ copies of a ($n+2$)-qubit
GHZ state;

(ii) $m$ qubits being distributed to each agent;

(iii) $m$ single-qubit Hadamard transformations and $m$ single-qubit
measurements being performed by each agent; and

(iv) $m$-bit classical message being sent to the receiver by each agent.

For the case of $m=1,$ the present method is not interesting since it
requires one more auxiliary qubit than the method in [13,14]. However, the
advantage of the present proposal appears when $m=2$ and becomes apparent as 
$m$ increases.

\begin{center}
{\bf IV. CONTROLLED TELEPORTATION OF MULTIPLE QUBIT-STRING INFORMATION TO
MANY DISTANT RECEIVERS }
\end{center}

It is interesting to note that the method described above can be further
extended to teleport multiple qubit-string information to many distant
receivers via the control of many agents in a network. Suppose that Alice
holds $k$ qubit strings labeled by $1,2,...,k$. The qubit string $l$
contains $m_l$ message qubits, which is initially in the state $%
\prod_{i=1}^{m_l}(\alpha _{i,l}\left| 0\right\rangle _{i,l}+\beta
_{i,l}\left| 1\right\rangle _{i,l})$ ($l=1,2,...,k$). The state of the $k$
qubit strings is given by $\prod_{l=1}^k\prod_{i=1}^{m_l}(\alpha
_{i,l}\left| 0\right\rangle _{i,l}+\beta _{i,l}\left| 1\right\rangle
_{i,l}). $ Now, Alice wishes to teleport the $k$ qubit-string information to 
$k$ distant receivers (the message carried by the qubit string $l$ is for
the receiver $l$) via the control of $n$ agents in a network, such that each
receiver can fully recover the original quantum message of the corresponding
qubit string only if all the agents cooperate. The present task can be
implemented with the following EPR-GHZ entangled state 
\begin{eqnarray}
&&\ \ \ \ \ \ \ \ \ \ \ \ \ \ \ \ \ \ \ \
\prod_{l=1}^k\prod_{i=1}^{m_l}(\left| 00\right\rangle _{i^{\prime }i^{\prime
\prime },\,l}+\left| 11\right\rangle _{i^{\prime }i^{\prime \prime
},\,l})\otimes \left| GHZ\right\rangle _{+}  \nonumber \\
&&\ \ \ \ \ \ \ \ \ \ \ \ \ \ \ \ \ \ \ \
+\prod_{l=1}^k\prod_{i=1}^{m_l}(\left| 00\right\rangle _{i^{\prime
}i^{\prime \prime },\,l}-\left| 11\right\rangle _{i^{\prime }i^{\prime
\prime },\,l})\otimes \left| GHZ\right\rangle _{-},
\end{eqnarray}
where $\left| GHZ\right\rangle _{\pm }$ are the GHZ states (17) and (18) of
the ($n+1$) GHZ qubits shared by the $n$ agents and Alice; the $m_l$ EPR
qubits ($1^{\prime \prime },2^{\prime \prime },...,m_l^{\prime \prime }$)
for the set $l$ belong to the receiver $l,$ while the other $m_l$ EPR qubits
($1^{\prime },2^{\prime },...,m_l^{\prime }$) for the set $l$ are kept by
Alice. Here, the set $l$ represents `` the qubit string $l$ and the $m_l$
EPR pairs shared by Alice and the receiver $l$ '' ($l=1,2,...k$). The state
of the whole system is thus given by 
\begin{eqnarray}
&&\ \ \ \ \ \ \ \ \ \ \ \ \ \prod_{l=1}^k\prod_{i=1}^{m_l}\left[ (\alpha
_{i,l}\left| 0\right\rangle _{i,l}+\beta _{i,l}\left| 1\right\rangle
_{i,l})(\left| 00\right\rangle _{i^{\prime }i^{\prime \prime },\,l}+\left|
11\right\rangle _{i^{\prime }i^{\prime \prime },\,l})\right] \otimes \left|
GHZ\right\rangle _{+}  \nonumber \\
&&\ \ \ \ \ \ \ \ \ \ \ \ \ +\prod_{l=1}^k\prod_{i=1}^{m_l}\left[ (\alpha
_{i,l}\left| 0\right\rangle _{i,l}+\beta _{i,l}\left| 1\right\rangle
_{i,l})(\left| 00\right\rangle _{i^{\prime }i^{\prime \prime },\,l}-\left|
11\right\rangle _{i^{\prime }i^{\prime \prime },\,l})\right] \otimes \left|
GHZ\right\rangle _{-}.
\end{eqnarray}
Now, Alice performs a series of two-qubit Bell-state measurements, which are
respectively on qubit pairs ($1,1^{\prime }$), ($2,2^{\prime }$),...($%
m_l,m_l^{\prime }$) for the set $l.$ After that, we obtain 
\begin{equation}
\ \ \ \ \prod_{l=1}^k\left| \psi \right\rangle _l\otimes \left|
GHZ\right\rangle _{+}+\ \prod_{l=1}^k\left| \psi ^{\prime }\right\rangle
_l\otimes \left| GHZ\right\rangle _{-},
\end{equation}
where $\left| \psi \right\rangle _l=\prod_{i=1}^{m_l}\left| \psi
\right\rangle _{i^{\prime \prime },l}$ and $\left| \psi ^{\prime
}\right\rangle _l=\prod_{i=1}^m\left| \psi ^{\prime }\right\rangle
_{i^{\prime \prime },l}$ are the states for the $m_l$ qubits ($1^{\prime
\prime },2^{\prime \prime },...,m_l^{\prime \prime }$) belonging to the
receiver $l$. $\left| \psi \right\rangle _{i^{\prime \prime },l}$ and $%
\left| \psi ^{\prime }\right\rangle _{i^{\prime \prime },l}$ are the states
of qubit $i^{\prime \prime }$ for the receiver $l,$ which depend on the
outcome of Alice's Bell-state measurement on the associated qubit pair ($%
i,i^{\prime }$) for the set $l,$ and take the form of (8) and (9),
respectively.

Note that the states $\left| \psi \right\rangle _l$ and $\left| \psi
^{\prime }\right\rangle _l$ have the same form as $\left| \psi \right\rangle 
$ and $\left| \psi ^{\prime }\right\rangle $ described in (7), respectively.
Therefore, based on (31) and using the above procedure, it is
straightforward to show that quantum information originally carried by each
qubit string can be recovered by the corresponding receiver, with the aid of
all the agents.

In order for each receiver to restore the original state of the
corresponding qubit string, the following procedure can be followed: (i)
Each agent and Alice need to perform a Hadamard transformation and then a
measurement on their respective GHZ qubits. (ii) Each agent and Alice need
to send each receiver their measurement results on their GHZ qubits. And
(iii) Alice needs to send the receiver $l$ the outcome of her Bell-state
measurements on the qubit pairs ($1,1^{\prime }$), ($2,2^{\prime }$),...($%
m_l,m_l^{\prime }$) for the set $l$, so that the receiver $l$ can recover
the original state of the qubit string $l.$

On the other hand, it can be shown from (31) that even if one agent does not
collaborate, the density operator for each qubit belonging to each receiver
takes the form of (15) or (16), i.e., no receiver can fully restore the
original state of the corresponding ``message qubit string'' without the
cooperation of all the agents.

To realize the present task, the present method requires only:

(i) $2\sum_{l=1}^km_l+n+1$ auxiliary qubits for preparing the state (29);

(ii) one qubit being assigned to each agent;

(iii) one single-qubit Hadamard transformation and one single-qubit
measurement being performed by each agent; and

(iv) one-bit classical message being sent to each receiver by each agent.

In contrast, to implement the same task, the method in [13,14] requires:

(i) $\sum_{l=1}^km_l(n+2)$ auxiliary qubits for preparing $\sum_{l=1}^km_l$
copies of a ($n+2$)-qubit GHZ state;

(ii) $\sum_{l=1}^km_l$ qubits being assigned to each agent;

(iii) $\sum_{l=1}^km_l$ single-qubit Hadamard transformations and $%
\sum_{l=1}^km_l$ single-qubit measurements being performed by each agent; and

(iv) $m_l$-bit classical message being sent to the receiver $l$ by each
agent.

One can see that even for $m_l=1$ and $k=2,$ the present method is
effective, since (a) the number of qubits distributed to each agent, the
number of Hadamard transformation by each agent, or the number of
measurement by each agent is $1$, which is, however, $2$ for the method in
[13,14]; and (b) the number of auxiliary qubits required is $n+5$, which is
smaller than $2n+4$ needed in the method in [13,14], when $n>1$. More
interestingly, with the increment of $m_l,k,$ or $n,$ the advantage of the
present method becomes very apparent.

\begin{center}
{\bf V. DISCUSSION AND CONCLUSION}
\end{center}

It should be pointed out that as far as the control efficiency of each agent
on teleportation, the present scheme is identical to those in [13,14]. This
is because the results (15) and (16) applied in the present proposal are the
same as those employed in [13,14]. However, as shown above, the present
scheme is extremely simple and economical in the realization of multi-qubit
quantum information teleportation via the control of many agents in a
network.

The scheme presented here works essentially through having
originally-non-entangled quantum informations, carried by any two message
qubits, to be entangled each other after Alice performs a series of
Bell-state measurements. This can be seen from equation (23). For example,
let us consider a $m=2$ case, i.e., teleporting the state $(\alpha _1\left|
0\right\rangle _1+\beta _1\left| 1\right\rangle _1)\otimes (\alpha _2\left|
0\right\rangle _2+\beta _2\left| 1\right\rangle _2$ of the two message
qubits ($1,2$) to Bob. Based on Eqs. (7)-(9), one sees that if Alice
measures the two qubits ($1,1^{\prime }$) and another two qubits ($%
2,2^{\prime }$) in the Bell states, e.g., $\left| \phi _{11^{\prime
}}^{+}\right\rangle $ and $\left| \phi _{22^{\prime }}^{+}\right\rangle ,$
the state (23) for the remaining qubit system will be 
\begin{eqnarray}
&&\ \ (\alpha _1\left| 0\right\rangle _{1^{\prime \prime }}+\beta _1\left|
1\right\rangle _{1^{\prime \prime }})(\alpha _2\left| 0\right\rangle
_{2^{\prime \prime }}+\beta _2\left| 1\right\rangle _{2^{\prime \prime
}})\otimes \left| GHZ\right\rangle _{+}\   \nonumber \\
&&\ \ +(\alpha _1\left| 0\right\rangle _{1^{\prime \prime }}-\beta \left|
1\right\rangle _{1^{\prime \prime }})(\alpha _2\left| 0\right\rangle
_{2^{\prime \prime }}-\beta _2\left| 1\right\rangle _{2^{\prime \prime
}})\otimes \left| GHZ\right\rangle _{-}.
\end{eqnarray}
The result (32) implies that if Bob measures the qubit $1^{\prime \prime }$
in the state $\alpha _1\left| 0\right\rangle +\beta _1\left| 1\right\rangle $%
, he can predict that his qubit $2^{\prime \prime }$ must be in the state $%
\alpha _2\left| 0\right\rangle +\beta _2\left| 1\right\rangle .$ On the
other hand, if Bob detects the qubit $1^{\prime \prime }$ in the state $%
\alpha _1\left| 0\right\rangle -\beta _1\left| 1\right\rangle $, he knows
that his qubit $2^{\prime \prime }$ must be in the state $\alpha _2\left|
0\right\rangle -\beta _2\left| 1\right\rangle .$ Therefore, after Alice
performs Bell-state measurements, the quantum informations originally
carried by the two message qubits ($1,2$) are not only transferred onto
Bob's qubits ($1^{\prime \prime },2^{\prime \prime }$) but also become
entangled each other.

In Refs. [13,14], single-qubit measurements are carried out in a new basis $%
\left| +\right\rangle =\left| 0\right\rangle +\left| 1\right\rangle $ and $%
\left| -\right\rangle =\left| 0\right\rangle -\left| 1\right\rangle .$ In
contrast, as shown above, the present single-qubit measurement is performed
in the basis $\left| 0\right\rangle $ and $\left| 1\right\rangle $. For
certain kinds of qubits (e.g., superconducting charge or flux qubits), it is
rather hard to make a measurement in the basis $\left| +\right\rangle $ and $%
\left| -\right\rangle $; but straightforward in the basis $\left|
0\right\rangle $ and $\left| 1\right\rangle .$ As a matter of fact, based on 
$\left| 0\right\rangle =\left| +\right\rangle +\left| -\right\rangle $ and $%
\left| 1\right\rangle =\left| +\right\rangle -\left| -\right\rangle ,$ it is
noted that Hadamard transformations are not necessary in the present
proposal, because the same results can be obtained when each agent performs
a measurement on his/her qubit in the basis $\left| +\right\rangle $ and $%
\left| -\right\rangle ,$ instead of a Hadamard transformation followed by a
measurement in the basis $\left| 0\right\rangle $ and $\left| 1\right\rangle
,$ and then sends his/her measurement result $\left| +\right\rangle $ or $%
\left| -\right\rangle $ (one-bit classical message) to the receiver(s).

Another point may need to be made here. As shown above, Alice's Bell-state
measurement, Alice's single-qubit operation (Hadamard
transformation/measurement), and each agent's operation are independently
performed on different qubits. Therefore, like the method in [13,14], the
present proposal actually does not require the operating order among Alice's
Bell-state measurement, Alice's single-qubit operation, and each agent's
operation.

Although we shall not attempt here a comprehensive study of the security of
the scheme against all possible forms of eavesdropping and/or cheating, we
believe that it is probably quite secure, for several reasons. First, the
eavesdropping by entangling ancillary qubits with Bob's qubits can be
revealed by comparing a subset of the states Bob received to the ones Alice
sent. Second, the qubits Alice sends to Bob are basically useless without
the classical information possessed by Alice. Hence, even if Eve was to
intercept the qubits intended for Bob, and replace them by fakes, and
somehow eavesdropped on the (classical) communication channels through which
all the agents disclose to Bob their measurement results, she would still
not be able to recover the message qubits' original states without access to
Alice's classical information (her measurement outcomes), given that Alice
sends her classical information to Bob using standard quantum cryptography
[14]. It is conceivable that an eavesdropper might obtain partial
information by entangling enough ancillary qubits with the qubits belonging
to all the agents and Bob, but presumably such entanglement could be
detected by tests conducted on `sample' EPR-GHZ entangled states initially
shared by Alice and the other parties.

In summary, we have presented a new method for teleporting multi-qubit
quantum information from a sender to a distant receiver, via the control of
many agents in a network. A special feature of our {\it entangling} quantum
information concept is that to implement a control of multi-qubit quantum
information teleportation, the present scheme needs to assign only one qubit
to each agent, followed by every agent performing only one Hadamard
transformation and one measurement, and then sending only one-bit classical
message to the receiver. As a result, the required auxiliary qubit
resources, the number of local operations, and the quantity of classical
communication are greatly reduced in the present proposal. The method
presented here can also be extended to implement a multi-party controlled
teleportation of multiple qubit-string quantum information to many distant
receivers. We believe that our scheme is of considerable interest,
especially because of its relatively straightforward nature in realizing
simultaneous control of multi-qubit quantum information teleportation in an
efficient and simple manner.

\begin{center}
{\bf ACKNOWLEDGMENTS}
\end{center}

CPY is very grateful to Julio Gea-Banacloche for the useful discussion and
Phyllis Bia for the help in the preparation of this manuscript. This
research was partially supported by National Science Foundation QuBIC
program (ECS-0201995), and AFOSR (F49620-01-1-0439), funded under the
Department of Defense University Research Initiative on Nanotechnology
(DURINT) Program and by the ARDA.


\begin{references}
\bibitem{s1}  C. H. Bennett, G. Brassard, C. Cr\'epeau, R. Jozsa, A. Peres,
and W. K. Wootters, Phys. Rev. Lett. {\bf 70,} 1895 (1993).

\bibitem{s2}  M. Fujii, Phys. Rev. A {\bf 68}, 050302 (2003).

\bibitem{s3}  N. Ba An, Phys. Rev. A {\bf 68}, 022321 (2003).

\bibitem{s4}  W. P. Bowen, N. Treps, B. C. Buchler, R. Schnabel, T. C.
Ralph, Hans-A. Bachor, T. Symul, and P. K. Lam, Phys. Rev. A {\bf 67},
032302 (2003).

\bibitem{s5}  T. J. Johnson, S. D. Bartlett, and B. C. Sanders, Phys. Rev. A 
{\bf 66}, 042326 (2002).

\bibitem{s6}  J. Fang, Y. Lin, S. Zhu, and X. Chen, Phys. Rev. A {\bf 67},
014305 (2003).

\bibitem{s7}  W. Son, J. Lee, M. S. Kim, and Y.-J. Park, Phys. Rev. A {\bf 64%
}, 064304 (2001).

\bibitem{s8}  E. F. Galv\~ao and L. Hardy, Phys. Rev. A {\bf 62}, 012309
(2000).

\bibitem{s9}  C.P.Yang and G. C. Guo, Chin. Phys. Lett. {\bf 16}, 628 (1999);%
{\it \ ibid.} {\bf 17}, 162 (2000).

\bibitem{s10}  D. Bouwmeester, J. W. Pan, K. Mattle, M. Eibl, H. Weinfurter,
and A. Zeilinger, Nature {\bf 390}, 575 (1997).

\bibitem{s11}  A. Furusawa, J. L. S\o rensen, S. L. Braunstein, C. A. Fuchs,
H. J. Kimble, and E. S. Polzik, Science {\bf 282}, 706 (1998).

\bibitem{s12}  M. A. Nielsen, E. Knill, R. Laflamme, Nature {\bf 396}, 52
(1998).

\bibitem{s13}  A. Karlsson and M. Bourennane, Phys. Rev. A {\bf 58}, 4394
(1998).

\bibitem{s14}  M. Hillery, V. Buzek, and A. Berthiaume, Phys. Rev. A {\bf 59}%
, 1829 (1999).

\bibitem{s15}  R. Cleve, D. Gottesman, and H. K. Lo , Phys. Rev. Lett. {\bf %
83}, 648 (1999).

\bibitem{s16}  A. Karlsson, M. Koashi, and N. Imoto, Phys. Rev. A {\bf 59},
162 (1999).

\bibitem{s17}  S. Bandyopadhyay, Phys. Rev. A {\bf 62}, 012308 (2000).

\bibitem{s18}  D. Gottesman, Phys. Rev. A {\bf 61}, 042311 (2000).

\bibitem{s19}  Li-Yi Hsu, Phys. Rev. A {\bf 68}, 022306 (2003).

\bibitem{s20}  A. C. A. Nascimento, J. M. Quade, and H. Imai, Phys. Rev. A 
{\bf 64}, 042311 (2001).

\bibitem{s21}  B. Aoun, M.Tarifi, e-print, quantum-ph/0401076.

\bibitem{s22}  E. Biham, B. Huttner, and T. Mor, Phys. Rev. A {\bf 54}, 2651
(1996); P. D. Townsend, Nature {\bf 385}, 47 (1997); S. Bose, V. Vedral, and
P. L. Knight, Phys. Rev. A {\bf 57}, 822 (1998).

\bibitem{s23}  J. W. Pan, M. Daniell, S. Gasparoni, G. Weihs, and A.
Zeilinger, Phys. Rev. Lett. {\bf 86}, 4435 (2001).

\bibitem{s24}  C. A. Sackett, D. Kielpinski, B. E. King, C. Langer, V.
Meyer, C. J. Myatt, M. Rowe, Q. A. Turchette, W. M. Itano, D. J. Wineland,
and C. Monroe, Nature {\bf 404}, 256 (2000).
\end{references}
\end{document}